\documentclass[prl,twocolumn,graphicx,amssymb,showpacs]{revtex4}
  \usepackage{color}
\usepackage{graphicx}
\begin{document}

\title{Reply to a Comment on ``Role of Potentials in the Aharonov-Bohm Effect''}

\author{Lev Vaidman}

\affiliation{ Raymond and Beverly Sackler School of Physics and Astronomy\\
 Tel-Aviv University, Tel-Aviv 69978, Israel}
\begin{abstract}
 Preceding Comment challenged my claim that  potentials might be just auxiliary mathematical tools and that they are not necessary for explaining physical phenomena. The Comment did not confront my explanation  without potentials of the Aharonov-Bohm effects appeared in the original paper, but stated that I cannot apply this explanation for seven other examples. In my reply I provide explanations using my method, of one of the examples, show that two other example are not relevant, and agree that the remaining examples require further analysis. However, I argue that none of the examples provide robust counter examples to my claim similar to the original Aharonov-Bohm setups which were explained in my Paper, so the Comment does not refute my claim.
      \end{abstract}

\pacs{03.65.-w, 03.65.Vf, 03.65.Ta, 03.65.Ud}

\maketitle

Quantum theory is about one century old, and it has only formulations based on potentials which,  due to the existence of local gauge transformations, do not have a local unique definition. This is in contrast with classical physics which has  a formulation with potentials, but also, alternatively,  with  uniquely defined fields locally interacting with particles. Only about a half century later,  Aharonov and Bohm (AB) \cite{AB} showed that it is not accidental by identifying  situations in which an electron  behaves differently in cases when   it is moving in field-free regions with different potentials. In my Paper \cite{myAB} I showed that considering the sources of electromagnetic field in  the AB setups quantum mechanically, the effect  can be explained using local field interactions and entanglement between the electron and the source. In their Comment, Aharonov, Cohen, and Rohrlich (ACR) \cite{Com}  argue  that my explanation  of the AB effect for the setups of the original AB paper is not capable  of explaining the AB effect in several other situations, and therefore, the consequence of the AB paper regarding the impossibility of explaining quantum effects via local actions of fields  still holds.

I am reluctant to accept nonlocality, unless there is an unambiguous clear proof for the opposite. The ACR arguments are not as clear as the AB argument which I was able to answer, so I am still optimistic regarding existence of a local explanation. Here I explain why I am not convinced by the ACR arguments.

The ACR ``prelude''   is probably the most serious challenge. An electron wave  packet passing through an interferometer with a constant magnetic field perpendicular to its plane  of motion  acquires the relative AB phase. It is not as dramatic as the original AB effect, since the electron moves in the region with nonzero magnetic field, but it demonstrates the same nonlocal feature, since   a local magnetic field on the paths of the electron wave packets does not provide an explanation for the AB phase. My Paper provides a calculation for the case in which the radius of the solenoid, $r$, is much smaller than the radius of the electron path, $R$. However, a more  complicated, but straightforward calculation shows that my mechanism explains the AB effect when $r$ is comparable to $R$ and even in the case $r>R$. My calculations were made in the framework of nonrelativistic quantum mechanics, so even if  $r>>R$, formally, the explanation holds.

If we are trying to extend this to the relativistic domain, there is a problem. If $r>>R$, and we  arrange to stop the cylinders at a particular time, we obtain a situation in which there is no current or charge density of the source, but a constant magnetic field inside the interferometer  continues to persist for a time $\frac{r-R}{c}$. We expect that in the AB experiment which takes less than this time, the effect will still take place, but without rotating cylinders my explanation will not hold. The analysis of this experiment in  relativistic domain is an important and  serious challenge.  I saw a recent attempt in this direction \cite{PR} but I  still do not see exactly how it works.  ACR also have not performed   an exact calculation, they have not showed that  such a calculation cannot lead to a local explanation.

 Their first example is described by Lagrangian (Eq. (1) in their Comment) which has a form of the Lagrangian of the Aharonov-Casher effect \cite{AC}.
 I have considered the AC effect and found that, as the AB effect, it also can be   explained locally \cite{PaACAB}. The difference in the ACR proposal is a particular vector potential, ${\bf  A} = {\bf \nabla} (\alpha \varphi_{12})$, which  they made up. It is not the vector potential of the electromagnetic interaction. My conjecture is that Nature is local and it does not have such interactions.

The second example just shows the strength of my approach because I was able to reproduce the effect ACR describe. The manifestation of the AB effect in their setup is the change in the energy spectrum of an electron as a function of encircled flux. (It can be obtained from the change of the Hamiltonian given by Eq.(5) of \cite{EPL}.) As noted by  ACR, the history has to be taken into account.  The interaction of the electron with the solenoid cannot be neglected and it invariably leads to entanglement between them. Therefore, one should consider the energy spectrum of the electron and the solenoid together. The change of the velocity of the cylinder which I found, Eq.~(6) of \cite{myAB}, leads to exactly the same change in the energy spectrum of the combined system.

The third example tells us that we need potentials if we want to describe the electron in a rotating frame. I never argued that potentials cannot describe physics. I argued that at present, after my local explanation of the AB effect,  we have no proof that without potentials there is no local way to describe physics. If there is an angular velocity of my frame (0 in our case) which allows a local explanation with fields, I will use this frame of reference. (I also do not see how the ACR successful description of the electron in  a rotating frame with potentials  proves the non-existence of other explanations.) Therefore, this example does not refute my conjecture.

The analysis of the AB experiment with a (super)conducting shield of the solenoid is a much more complicated task. It is plausible that the shield prevents accumulation of the phase by the cylinders, which is my explanation of the AB effect in the original setup, but now there are charges in the shield which might accumulate the phase. The actual experiment with  a superconducting shield  \cite{Tono} showed the AB effect with a particular value of elementary charges of the shield, $2e$. My model for explaining the AB effect is not flexible enough to take quantization of charge in the conducting shield into account. Currently I do not know how to perform such an analysis. But I also do not see that such analysis will refute my conjecture.    I encourage performing exact  analyses of these cases.

The ACR  sixth example considers  a velocity  measurement of an extended wave packet  of an electron passing near the solenoid. It is not very clear how this velocity measurement can be  performed, and the entanglement of such an electron with the solenoid complicates the matter even more. However,  the same point can be analyzed considering a superposition of two localized well separated wave packets of the electron passing near the solenoid at the moment that one packet passed the solenoid and another did not. Then, according to my approach,  a relative phase due to rotation of the cylinder (my model of the solenoid) is created, but ``it is not seen in the experiment''.

How  can the relative phase be observed? A natural way is to observe interference between the wave packets. But this requires bringing the wave packets together and then the cylinder makes different rotations again such that the phase disappears. One might try to observe the phase using local interactions of the wave packets with a probe particle in a superposition \cite{AVnl}. But then, the probe wave packets also acquire relative phase due to their interaction with the solenoid, so again, the phase will not be observed. Thus, I have a good explanation why the relative phase created by my mechanism has not been seen in an experiment.

Needless to say, I am not ready to accept the nonlocal action of fields, the alternative to potentials which ACR suggested  in their conclusion. I want to believe that there is a local account for everything in physics. My Paper removed the apparent counter example, the original AB effect. The ACR Comment poses interesting questions, but it does not provide a counter example of a similar strength.

  This work has been supported in part by the  Israel Science Foundation  Grant No. 1311/14.

\end{document}